\documentclass[12pt]{iopart}

\usepackage{
	amssymb, 
	amsthm, 
	graphicx, 
	xspace,
	overpic,
	ifthen,
	verbatim, 
	subfigure, 
	appendix,
	ifthen,
	color 
} 

\newcommand{\bra}[1]{\langle #1 \vert  }
\newcommand{\ket}[1]{\vert #1 \rangle  }

\newcommand{\braket}[2]{  \langle #1 \vert #2 \rangle  }

\newcommand{\matrixelement}[3]{  \langle #1 \vert #2 \vert #3\rangle  }

\newcommand{\ketbra}[2]{  \vert #1 \rangle \langle #2 \vert }

\newcommand{\projector}[1]{  \vert #1 \rangle \langle #1 \vert }

\newcommand{\Abs}[1]{\left| #1 \right|}

\newcommand{\dd}{\,\mathrm{d}} 

\newcommand{\MI}{\ensuremath{\mathcal{I}}}
\newcommand{\Sys}{\ensuremath{\mathcal{S}}}
\newcommand{\Env}{\ensuremath{\mathcal{E}}}
\newcommand{\Frag}{\ensuremath{\mathcal{F}}}

\newcommand{\newtext}{\color{black}}
\newcommand{\draftmode}{1}
\newcommand{\oldtext}{\ifthenelse{\equal{\draftmode}{1}}{\color[gray]{0.5}}{\color{black}}}
\newcommand{\notetoself}[1]{\ifthenelse{\equal{\draftmode}{1}}{{\color[rgb]{0,0,0.8} [#1]}}{}}

\usepackage{setstack}
\newcommand{\operatorname}[1]{\mathop{\mathrm{#1}}}
\newcommand{\eqref}[1]{(\ref{#1})}
\newcommand{\binom}[2]{ \left( \begin{array}{c} #1 \\ #2 \end{array} \right) }

\newcommand{\Eqref}[1]{\eref{#1}}
\newcommand{\Figref}[1]{figure \ref{#1}}

\newcommand{\pbwidthfactor}{0.7} 
\newcommand{\dualwidthfactor}{0.48} 
\newcommand{\widewidthfactor}{0.96} 

\newcommand{\bbb} {\ensuremath{\mathbb{B} }}
\newcommand{\sss}{\ensuremath{\mathbb{S}}}
\newcommand{\bbbb}{{\overline{\bbb}}}
\newcommand{\sone}{S_{ \vec{x}_1}}
\newcommand{\stwo}{S_{ \vec{x}_2}}
\newcommand{\sonetwo}[1]{S_{ \vec{x}_1}^{#1}{}^\dagger S_{ \vec{x}_2}^{#1}}
\newcommand{\stwoone}[1]{S_{ \vec{x}_2}^{#1}{}^\dagger S_{ \vec{x}_1}^{#1}}

\renewcommand{\draftmode}{0} 

\begin{document}


\title[Redundant Information from Thermal Illumination]{Redundant Information from Thermal Illumination: Quantum Darwinism in Scattered Photons}
\date{\today}
\author{C.~Jess~Riedel$^{1,2}$ and Wojciech~H.~Zurek$^{1,3}$}
\address{$^1$Theory Division, LANL, Los Alamos, New Mexico 87545, USA}
\address{$^2$Department of Physics, University of California, Santa Barbara, California 93106, USA}
\address{$^3$Santa Fe Institute, Santa Fe, New Mexico 87501, USA}
\eads{criedel@physics.ucsb.edu}


\newtext 
\begin{abstract}
We study quantum Darwinism, the redundant recording of information about the preferred states of a decohering system by its environment, for an object illuminated by a blackbody.  We calculate the quantum mutual information between the object and its photon environment for blackbodies that cover an arbitrary section of the sky. In particular, we demonstrate that more extended sources have a reduced ability to create redundant information about the system, in agreement with previous evidence that initial mixedness of an environment slows---but does not stop---the production of records.  We also show that the qualitative results are robust for more general initial states of the system.
\end{abstract}


\maketitle

\section{Introduction}
\label{sec:introduction}

\newtext

The theory of decoherence \cite{SchlosshauerText,JoosText,Zurek2003} is supported by striking experimental evidence \cite{Brune1996, Hornberger2003b, Hasselbach2007} and helps explain the emergence of the classical realm in a purely quantum universe.  Classicality is marked by several characteristics that, at first sight, appear not to be native to quantum mechanics.  The progress over the past three decades has been due to the realization that these classical aspects of our Universe can arise dynamically, and are sensitive to the details of the systems and interactions being studied.

Decoherence can explain why effectively classical  \emph{pointer} states (e.g. Gaussian wavepackets) are preferred for certain systems, in the sense that the system's pointer state is unaffected by its interaction with the environment, so that any other initial state will quickly become approximately diagonal in the associated pointer basis.  It also explains why (and under what circumstances) interference between components in a system-environment entangled state can be ignored, the environment can be traced out, and the system state can be regarded as having ``reduced'' to an approximate mixture of pointer states.

Nevertheless, there are still deep unanswered questions about the quantum-classical transition.  In particular, decoherence alone does not explain two aspects of classical mechanics which we take for granted: the states of classical systems are \emph{robust} and \emph{objective}.  By ``robust'', we mean that observers may discover an initially unknown state of the classical system without disturbing it.  By ``objective'', we mean that multiple observers may independently find out the state of the same classical system, that they will all agree on the answer, and that their measurements will leave the system in the preexisting (objective) state.

In general, quantum states are neither robust nor objective even after decoherence has eliminated obviously quantum superpositions.  When an observer measures a system in something other than it's pointer basis, he dramatically affects the state by re-preparing the system in an eigenstate of the observable he has measured.  Therefore, when multiple observers each measure a system in different bases, they will each get different, incompatible results, and the last measurement will leave the system in a state that has little to do with the states revealed by its predecessors.  What, then, explains the objective and robust nature of classical macroscopic objects?

Quantum Darwinism provides the answer to this question \cite{Zurek2000, Zurek2009}.  It recognizes that real observers do not typically interact \emph{directly} with the system they measure.  Instead, the system is immersed in and correlated with a (decohering) environment. The observer then interacts with a fraction of that environment to find out its states -- to become correlated with the system -- measuring it indirectly.  For instance, when we ``measure'' the position of a chair by looking at it, our eyes do not directly interact with the chair.  The chair's state is not affected by whether or not we open our eyes.  By opening our eyes, we merely allow them (and hence, our neurons) to become correlated with some of the photons scattered by chair (and hence, its position).

The environment in the Universe we inhabit acts as an information channel through which the observer finds out about the system \cite{Ollivier2005,Zurek2009}. But this information is filtered, as only the observables that are recorded in many copies in the environment can be found out from the intercepted fragment. This means that observers cannot choose any arbitrary basis in which to measure; they are restricted in the type of information that can be acquired about the system by the nature of the system-environment interaction.  Under the condition of effective decoherence, this information can only describe the pointer states of the system, \emph{not} superpositions thereof \cite{Ollivier2004,Ollivier2005}.  Indeed, the no-cloning theorem \cite{Wootters1982, Dieks1982} implies that arbitrary quantum states of the system will not be able to proliferate in this manner, as only certain preferred states (that turn out to be pointer states) can be imprinted onto many fragments of the environment \cite{Zurek2007a}.  In fact, observables complementary to the pointer states effectively become inaccessible after decoherence has set in since they can only be recovered through \emph{global} measurements on the whole of the environment.

So based only on the assumption that typical observers learn about the system through the environment (i.e. independent of any details about the size of the observers or the way they interact with the environment), we can conclude that (1) observers do not disturb the system (``robustness'') and (2) all observers can learn \emph{only} about the pointer basis and, consequently, will agree (``objectivity'').

This description, which is related \cite{Dalvit2001, Dziarmaga2004} to the formalism of quantum trajectories \cite{CarmichaelText}, can only hold when multiple observers can actually determine the state of the system by sampling just part of the environment.  That is, information about the system must be recorded \emph{redundantly} in the environment.  This will not always be the true.  In some cases of effective decoherence, the environment simply does not make multiple copies of the information (e.g. collisional decoherence from a single high-energy environmental particle).  In other cases, strong self-interactions scramble correlations and prevent information about the system from being extracted from any accessible fraction of the environment (e.g. collisional decoherence from air molecules). 

The capacity of quantum Darwinism to explain the quantum-classical transition then rests on whether decoherence in everyday settings actually induces sufficient redundancy such that the classical approximations of robustness and objectivity are justified.  This has been investigated for a spin\nobreakdashes-$\frac{1}{2}$ particle monitored by a pure \cite{Blume-Kohout2005} and mixed \cite{Zwolak2009a,Zwolak2010} bath of spins and a harmonic oscillator monitored by a pure bath of oscillators \cite{Blume-Kohout2008,Paz2009}.  We recently showed for the first time that a physically realistic setting---an object illuminated by a point-source blackbody---does in fact lead to enormous redundancies \cite{Riedel2010}.  In this work, we generalize that analysis to include partially-angularly-mixed illumination and arbitrary initial object wavefunctions.  We confirm evidence in earlier studies \cite{Zwolak2009a,Zwolak2010} that initial partial mixedness of the environment hampers---but does not eliminate---its ability to redundantly record the state of the system.  We also show that quantum Darwinism in our model is robust for general initial states of the system.

\section{Quantum Darwinism in Photon Collisional Decoherence}
\label{sec:stage}

When an object in a mesoscopic superposition is exposed to radiation, scattering photons will quickly reduce its pure, nonlocal state to a mixture of localized alternatives via collisional decoherence \cite{Joos1985}.  (See also \cite{Diosi1995,Gallis1990,Hornberger2003a,Hornberger2006} for refinements and corrections.) The fantastic rate of collisional decoherence has been confirmed experimentally \cite{Kokorowski2001,Uys2005}.

Observers typically access a small part of the environment (in this case, the photons that enter one's eye), so we will estimate how much information about the object is available in a subset of the environmental photons.  For simplicity, we assume our environment consists of a large but fixed number $N$ of identical photons: $\Env = \bigotimes_{n = 1}^N \Env_i$, where $\Env_i$ is the Hilbert space of a single photon in a box of volume $V$.  We then define $\Frag_f = \bigotimes_{n = 1}^{fN} \Env_i$ to be the \emph{fragment} corresponding to some fraction $f$ of the environment composed of $f N$ photons.  Since each photon has the same initial conditions and interactions, the choice of photons with which to construct the fragment is unimportant.  To get our final results, we will take $V$ and $N$ to infinity while holding the physical photon density $N/V$ constant.

The primary quantity investigated will be the quantum mutual information 
\begin{eqnarray}
\MI_{\Sys : \Frag} = H_{\Sys} + H_{\Frag} - H_{\Sys, \Frag}
\end{eqnarray}
between the system $\Sys$ and fragment $\Frag$, where $H$ denotes the von Neumann entropy.  From this we will calculate the \emph{redundancy} $R_{\delta}$, which is the number of distinct fragments in the environment that supply, up to an \emph{information deficit} $\delta$, the classical information about the state of the system.  More precisely, $R_{\delta} = 1/f_\delta$, where $f_\delta$ is the smallest fragment such that $\MI_{\Sys : \Frag_{f_\delta}} = (1-\delta)\overline{H_\Sys}$.  ($\overline{H_\Sys}$ is the maximum entropy of $\Sys$.  Only very large fragments $f \geq 0.5$ will be able to have perfect classical information, $\MI_{\Sys : \Frag_{f_\delta}} = \overline{H_\Sys}$, about the object \cite{Blume-Kohout2005}.)  At any given time, the redundancy is the measure of objectivity; it counts the number of observers who could each independently determine the state of the system (up to a small residual uncertainty $\delta$) by interacting with disjoint fragments of the environment.

\newtext


\newtext


Following Joos and Zeh \cite{Joos1985}, we take our system to be a dielectric sphere of radius $a$ and relative permittivity $\epsilon$ in a pure state.
The system and the photons in the environment are assumed to be initially unentangled: $\rho^0 = \rho_{\Sys}^0~\otimes~\rho_{e}^0~\otimes~\cdots~\otimes~\rho_{e}^0$, where $\rho_{\Sys}$ and $\rho_{e}$ are the density matrices of the system and of a single photon, respectively, and a superscript ``$0$'' denotes prescattering states.  

\begin{figure} [bt!]
  \centering 
  \includegraphics[width=\widewidthfactor \columnwidth ]{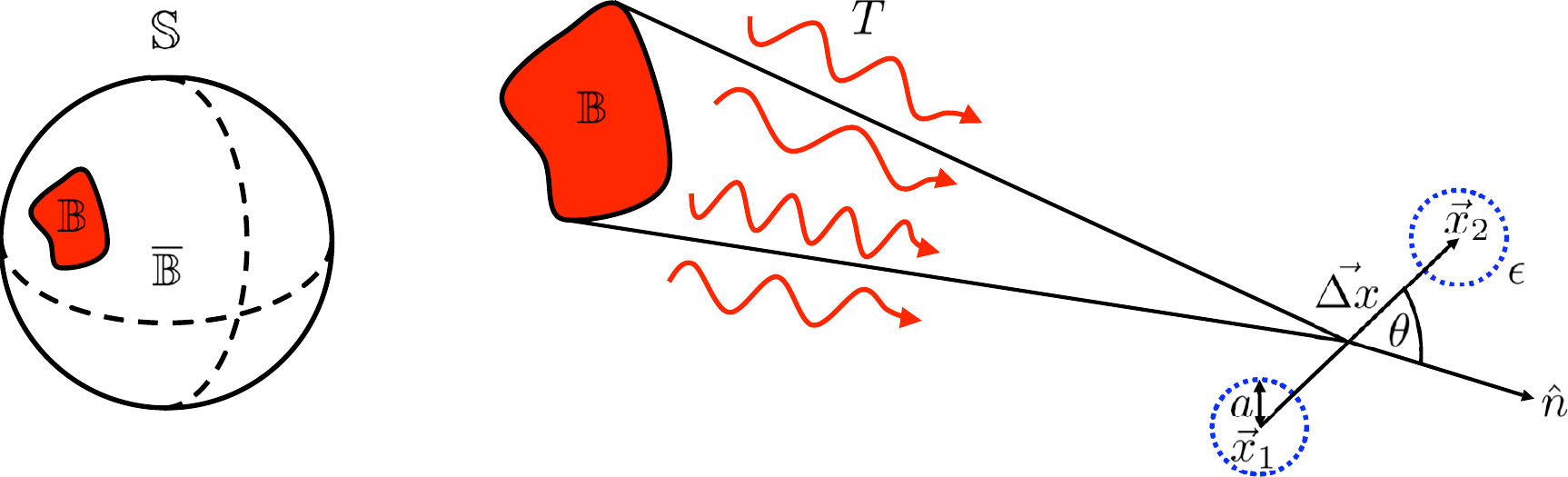}
  \caption{A dielecric sphere of radius $a$ and permittivity $\epsilon$ is initially in a superposition with separation $\Delta x = |\vec{x}_1-\vec{x}_2|$.  The object is subjected to radiation from a blackbody at temperature $T$ that originates from a patch $\bbb \subset \sss$ (where $\sss$ is the unit sphere ``sky'' as seen from $\Sys$) with solid angle $\Omega \le 4\pi$.  The complement of $\bbb$ is $\bbbb$.  The photons propagate in directions labeled by $\hat{n}$, which makes an angle $\theta$ with the vector $\vec{\Delta x}$.}
  \label{fig:intro}
\end{figure}

The system is illuminated by photons originating from a far away blackbody of temperature $T$ that covers $\bbb \subset \sss$, where $\sss$ is the unit sphere ``sky'' as seen from $\Sys$.  Let $\Omega \le 4\pi$ be the solid angle measure of $\bbb$. See \fref{fig:intro}.

In the Hilbert space of a single photon, we break the momentum eigenstates into a tensor product $\ket{\vec{k}} = \ket{k_i}\ket{\hat{n}}/k$ of magnitude and directional eigenstates.  (The factor of $1/k$ comes from the Jacobian determinant associated with fact that the kets in this infinite-dimensional Hilbert space are densities, not normalized vectors.)   Blackbody illumination is then described by 
\begin{eqnarray}
\rho_{e}^0 = \int_0^{\infty} \dd k \, p(k) \projector{k} \otimes \int_\bbb \frac{\dd \hat{n}}{\Omega} \projector{\hat{n}} .
\end{eqnarray}
where $p(k) \propto  k^2/[\exp(k c / k_B T)-1]$ and $c$ is the speed of light.  Above, the probability distribution over the solid angle is the normalized characteristic function of $\bbb$. This ``step-function''  illumination is not contrived. Perfect blackbodies are Lambertian, which means that surface elements appear to have the same brightness no matter the angle of viewing.  In other words, the sun appears as a uniform disk of illumination in the sky; it is not dimmer near the edge.


We ignore the self-Hamiltonian of the object and assume it is heavy enough to have negligible recoil from photon scattering, so the evolution is governed by
\begin{eqnarray}
U_t \ket{\vec{x}}\ket{\vec{k}} = \ket{\vec{x}} S_{\vec{x}} \ket{\vec{k}},
\end{eqnarray}
where $S_{\vec{x}}$ is a scattering matrix acting on the single photon state when the particle is located at $\vec{x}$.  Elastic scattering leads to
\begin{eqnarray}
\label{eq:elastic-scatt}
S_{\vec{x}} \ket{\vec{k}} = S_{\vec{x}} \ket{k} \ket{\hat{n}} /k = \ket{k} S_{\vec{x}}^{k} \ket{\hat{n}}/k \, ,
\end{eqnarray}
under the magnitude-direction decomposition of the photon momentum states.

\subsection{Decoherence}

For now, we take the initial state of the object to be a (Schr\"odinger) ``cat'' state: $\rho_\Sys^0 = \projector{\psi}$, $\ket{\psi} = (\ket{\vec{x}_1} + \ket{\vec{x}_2})/\sqrt{2}$.  The decoherence of the superposition is governed by the decay of the off-diagonal terms in the position basis,
\begin{eqnarray}
\label{eq:position-decoh}
\left|\matrixelement{ \vec{x}_1 }{ \rho_\Sys }{ \vec{x}_2}\right|^2 = \gamma^N \left|\matrixelement{ \vec{x}_1 }{ \rho_{\Sys}^0 }{ \vec{x}_2 }\right|^2 ,
\end{eqnarray}
where 
\begin{eqnarray}
\label{eq:gamma}
\gamma  \equiv  \left|\operatorname{Tr} \left[ \stwo \rho_{e}^0 \sone^\dagger \right] \right|^2 .
\end{eqnarray}
The complex value $\gamma$ is the \emph{decoherence factor} attributable to a single photon.  The two-dimensional $\rho_\Sys$ can be diagonalized and its entropy is
\begin{eqnarray}
\label{eq:H_S-open}
H_{\Sys} &=& \ln 2 - \sum_{n=1}^{\infty} \frac{\Gamma^{n}}{2n(2n-1)} \\
\label{eq:H_S-closed}
&=& \ln 2 - \sqrt{\Gamma} \operatorname{arctanh}{\sqrt{\Gamma}} - \ln \sqrt{1-\Gamma}
\end{eqnarray}
where $\Gamma \equiv \gamma^{N}$ is the decoherence factor associated with the environment as a whole.  

The function 
\begin{eqnarray}
\label{eq:h-func}
h(x) = \sqrt{x} \operatorname{arctanh}{\sqrt{x}} + \ln \sqrt{1-x} = \sum_{n=1}^\infty \frac{x^n}{2n(2n-1)}
\end{eqnarray}
will appear often.  Note that $h(0) = 0$, $h(1) = \ln 2$, and
\begin{eqnarray}
\label{eq:h-bounds}
x/2 \le h(x) \le x \ln 2 .
\end{eqnarray}
Also, $h(x)$ is analytic and monotonic on the interval $[0,1]$ and, for small $x$, $h(x) \approx x/2$.

Applying \Eqref{eq:elastic-scatt} gives
\begin{eqnarray}
\label{eq:gamma-hat}
\operatorname{Tr} \left[ \sone \rho_{e}^0 \stwo^\dagger \right] = \int_0^{\infty} \dd k \, p(k)  \int_\bbb \frac{\dd \hat{n}}{\Omega} \matrixelement{n} {\stwo^{k \dagger} \sone^{k}} {n} .
\end{eqnarray}
To get the key matrix element appearing on the r.h.s., we use the classical cross section of a dielectric sphere \cite{JacksonEMcrosssection} in the dipole approximation ($\lambda \gg a$) and assume the photons are not sufficiently energetic to resolve the superposition individually ($\lambda \gg \Delta x$) \footnote{High energy photons are uninteresting as they can easily ``see'' the separation $\Delta x$ and each will become completely correlated with the position of the object; the redundancy will be just be equal to the number of scattered photons.}. This gives \cite{Riedel2010}
\begin{eqnarray}
\label{eq:key-matrix-element}
\bra{n} \stwo^{k \dagger} \sone^k \ket{n} =  1-\frac{1}{V} \frac{2 \, \pi}{15} (3+11 \cos^2 \theta) \frac{\tilde{a}^6 \Delta x^2 t k^6 c}{\hbar^6} + O\left(V^{-2}\right) ,
\end{eqnarray}
where $\tilde{a} \equiv a [(\epsilon-1)/(\epsilon-2)]^{1/3}$ is the effective radius of the object, $\theta$ is the angle between $\hat{n}$ and $\vec{\Delta x}$, and $t$ is the elapsed time.  

For increasing $V$, photon momentum eigenstates become diffuse so individual photons decohere the state less and less.  In other words, $\gamma \rightarrow 1$ because $S_{\vec{x}}^k \to I$.  This is balanced by an increasing number of photons in the box. In the limit $V,N \to \infty$, we combine \Eqref{eq:gamma}, \Eqref{eq:gamma-hat}, and \Eqref{eq:key-matrix-element} and use $e = \lim_{q \to \infty} (1 + 1/q)^q$ to get the decoherence factor
\begin{eqnarray}
\Gamma = \lim_{V,N \to \infty} \gamma^{N} =  e^{- t / \tau_D},
\end{eqnarray}
where $\tau_D$ is the \emph{decoherence time}.  It's inverse is the decoherence \emph{rate} \footnote{This is related to Joos and Zeh's convention by $\tau_D^{-1} = 2 \tau^{-1} = 2 \Lambda \Delta x^2$, where $\tau$ and $\Lambda$ are the ``characteristic time'' and ``localization rate'' of \cite{Joos1985}.}, 
\begin{eqnarray}
\label{eq:universal-decoh-rate}
\tau_D^{-1} &=& \frac{N}{V} \left( \frac{4 \pi}{15} \right) \left( \frac{8! \zeta(9)}{2! \zeta(3)} \right) \frac{\tilde{a}^6 \Delta x^2 k_B^6 T^6}{c^5 \hbar^6} \int_\bbb \frac{\dd \hat{n}}{\Omega} \left( 3 + 11 \cos^2 \theta \right) \\
&=& \Omega \left( \frac{8! \zeta(9)}{15 \pi^2} \right) \frac{\tilde{a}^6 \Delta x^2 k_B^9 T^9}{c^8 \hbar^9} \left\langle 3 + 11 \cos^2 \theta  \right\rangle_{\bbb} ,
\end{eqnarray}
where $\zeta(n)$ denotes the Riemann zeta function, which arises from integrating over the thermal distribution.  In the second line, we have expressed the number density of blackbody radiation in terms of the apparent solid angle, $N/V = \Omega \zeta(3) k_B T / (2 \pi^3 c^3 \hbar^3)$, and we have rewritten the integral to emphasize that it is merely the average of the trigonometric quantity $3+11 \cos^2 \theta$ over $\bbb$.  If we ignore the order-unity change in that average (it is obviously constrained to lie between $3$ and $14$) then \emph{the decoherence rate is linear with the apparent size of the blackbody}.  This is natural because thermal radiation is uncorrelated. Each new photon contributes an independent multiplicative decoherence factor, which combine additively in the decoherence rate.

For blackbodies that are far enough away to be approximated as point sources, the \emph{irradiance} $I$ (radiative power per unit area) is a more physically accessible quantity than the the solid angle $\Omega$, especially in the presence of optical distortion.  In that case we use $N/V = (I / c\, k_B T)[2! \zeta(3)]/[3! \zeta(4)]$, to get the point-source decoherence rate \cite{Riedel2010}
\begin{eqnarray}
\tau_D^{-1} = \left(\frac{4\pi}{15} \right) \left(\frac{8! \zeta(9)}{3! \zeta(4)} \right) (3+11 \cos^2 \theta) \frac{ I \tilde{a}^6 \Delta x^2 k_B^5 T^5}{c^6 \hbar^6}.
\end{eqnarray}
where $\theta$ is the angle between $\vec{\Delta x}$ and the point source.

At the opposite extreme where $\Omega = 4\pi, \bbb = \sss$, we recover the decoherence rate for isotropic thermal illumination \cite{Hornberger2003a}: 
\begin{eqnarray}
T_D^{-1} = \left( 16\frac{8! \zeta(9) }{9 \pi} \right) \frac{\tilde{a}^6 \Delta x^2 k_B^9 T^9}{c^8 \hbar^9} .
\end{eqnarray}
This is the decoherence rate when the system is surrounded by a uniform blackbody, e.g. inside an oven.  For the corresponding decoherence time, we have introduced the symbol $T_D \equiv \tau_D|_{\bbb = \sss}$ since it will serve as a useful $\bbb$-independent timescale in the rest of this paper.

For concreteness, consider a blackbody that appears as a disk on the sky centered on a direction $\hat{z}$ which makes an angle $\chi$ with $\vec{\Delta x}$.  That is, $\bbb = \{ (\theta, \phi) \in \sss \mid \theta \le \theta_0\}$ for some maximum polar angle $\theta_0$.  See \fref{fig:disk-alpha}.  For such disks, the decoherence rate is  
\begin{eqnarray}
\label{eq:norm-decoh-rate-disk}
\fl \tau_D^{-1} =  \frac{1}{80} \left[40 - \cos \theta_0 (51 - 33 \cos^2 \chi) +  \cos^3 \theta_0 (11 -  33 \cos^2 \chi) \right] T_D^{-1} .
\end{eqnarray}
This is plotted in \fref{fig:disk-alpha}(a) in terms of the solid angle $\Omega = 2 \pi(1-\cos \theta_0)$.  It is monotonically increasing with $\Omega$ since each additional photon can only further decohere the system.

\subsection{Quantum Darwinism}

To get the redundancy in the environment, we will need to find the mutual information $\MI_{\Sys : \Frag} = H_{\Sys} + H_{\Frag} - H_{\Sys \Frag}$.  We can avoid calculating $H_{\Sys \Frag}$ by using the identity [Eq.~(8) of \cite{Zwolak2010}]
\begin{eqnarray}
\label{eq:mi-identity}
\MI_{\Sys : \Frag} = \left[H_{\Frag} - H_{\Frag}^0 \right] + \left[ H_{\Sys d \Env} - H_{\Sys d \Env / \Frag} \right] ,
\end{eqnarray}
where $H_{\Sys d \Env} = H_{\Sys}$ is the entropy of the system as decohered by the entire environment $\Env$, and $H_{\Sys d \Env / \Frag}$ is the entropy of the system if it were decohered by only $\Env/\Frag$.  We get $H_{\Sys d \Env / \Frag}$ from $H_{\Sys}$, \Eqref{eq:H_S-open}, by making the replacement $\Gamma \to \Gamma^{1-f}$. 

The calculation of $H_{\Frag}$ is tedious, and we relegate the details to the appendix.  The change in the entropy of the fragment $\Frag$ is
\begin{eqnarray}
\label{eq:delta-H_F}
\Delta H_{\hat{\Frag}} \equiv H_{\hat{\Frag}} - H_{\hat{\Frag}}^0 =  \ln 2 - \sum_{m=1}^{\infty} \frac{\Gamma^{m \alpha f}}{2m(2m-1)} ,
\end{eqnarray}
where the effect of the initial mixedness of the environment on the production of records is accounted for by the single parameter
\begin{eqnarray}
\label{eq:general-alpha}
\alpha = \frac{\int_\bbb \dd \hat{n} \int_{\bbbb} \dd \hat{m}  |g(\hat{n},\hat{m})|^2 }{\int_\bbb \dd \hat{n} \int_\sss \dd \hat{m} |g(\hat{n},\hat{m})|^2} .
\end{eqnarray}
For reasons that we explain below, we call $\alpha$ the \emph{receptivity} of the environment with respect to the decoherence process.  Above, $\bbbb \equiv \sss \backslash \bbb$ is the complement of $\bbb$ and we define  $\Abs{g(\hat{n},\hat{m})}^2 \equiv \int \dd k p(k) \Abs{\matrixelement{\hat{n}}{(\sonetwo{k}-I)}{\hat{m}}}^2$.  The function $\Abs{g(\hat{n},\hat{m})}^2$ is a measure of the distinguishability of the out states $\sone \ket{\hat{n}}$ and $\stwo \ket{\hat{m}}$ for different incoming angles $\hat{n}$ and $\hat{m}$ of $\Env$ \emph{and} for different locations $\vec{x}_1$ and $\vec{x}_2$ of $\Sys$.  The receptivity $\alpha$ is a dimensionless ratio constructed from this function that, from the form of \Eqref{eq:general-alpha}, we know obeys $0 \le \alpha \le 1$.  

\begin{figure} [bt!]
  \centering
  \includegraphics[width=\pbwidthfactor \columnwidth ]{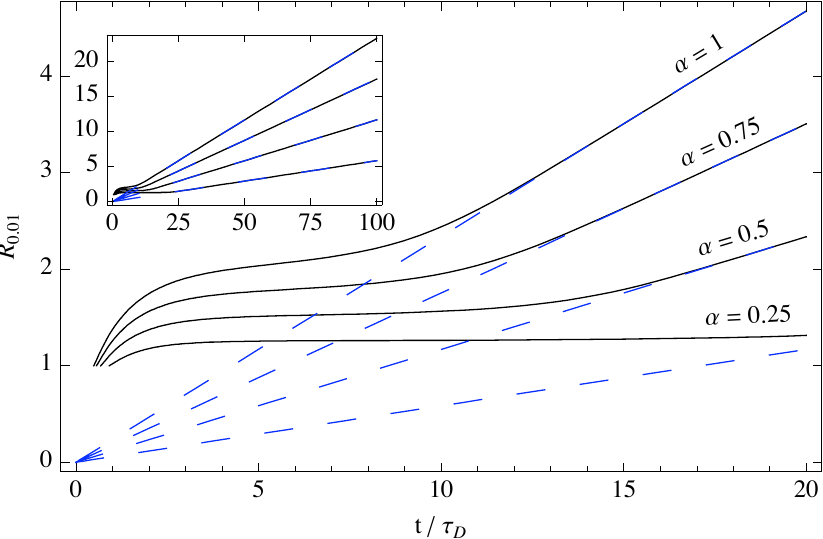}
  \caption{The redundancy $R_\delta$ versus time for an object illuminated by blackbody radiation of varying apparent solid angle, with information deficit $\delta = 0.01$.  The exact redundancy (solid black lines) is plotted starting when $R_\delta = 1$, i.e. when it first become possible for an observer to determine the state of the system to within the information deficit $\delta$ by accessing the entire environment.  For times $t \gg \tau_D$, the redundancy is well approximated by the linear estimate  (dashed blue line) in \Eqref{eq:redundancy}.}
  \label{fig:alpha-redun}
\end{figure}

Now that we have the change in fragment entropy $\Delta H_{\hat{\Frag}} = H_{\hat{\Frag}} - H_{\hat{\Frag}}^0$, we use \Eqref{eq:mi-identity} to finally write down the mutual information
\begin{eqnarray}
\label{eq:universal-MI-open}
\MI_{\Sys : \Frag} =  \ln 2 + \sum_{m=1}^{\infty} \frac{\Gamma^{(1-f) m} - \Gamma^{\alpha f m} - \Gamma^{m}}{2m(2m-1)} .
\end{eqnarray}
The summations in \Eqref{eq:universal-MI-open} can be written in a closed form analogous to \Eqref{eq:H_S-closed}, but the power series is more useful for calculating the redundancy.  For large times, $\Gamma = \exp(- t / \tau_D)$ is exponentially small and the sum is dominated by the lowest power of $\Gamma$.  If $0<f<1/2$ and $\alpha \neq 0$, then $0 < \alpha f < (1-f) < 1$ and
\begin{eqnarray}
\label{eq:approx-MI}
\MI_{\Sys : \Frag_f} &\approx& \ln 2 - \frac{1}{2} \Gamma^{\alpha f} . 
\end{eqnarray}
So long as the information deficit is not unreasonably large, $\delta < 1/(2 \ln 2) \approx 0.72$, we can estimate the redundancy in the limit $t \gg \tau_D$:
\begin{eqnarray}
\label{eq:redundancy}
R_\delta &\approx& \frac{t /\tau_R}{\ln [(2\, \delta \ln 2)^{-1}]} 
\end{eqnarray}
where
\begin{eqnarray}
\tau_R^{-1} = \alpha \, \tau_D^{-1} 
\end{eqnarray}
is the \emph{redundancy rate}---the characteristic rate at which records about the state of the system are produced \footnote{The times for which this is a good approximation to the true redundancy are shown in \fref{fig:alpha-redun}.  For added rigor, we can use \Eqref{eq:h-bounds} to get $\MI_{\Sys : \Frag} > \ln 2 (1  - \Gamma^{\alpha f} - \Gamma)$, which, for $t > \tau_d \ln (2/\delta)$, yields this  lower bound on the redundancy: $R_\delta > (t /\tau_D)/ \ln [( \delta -\Gamma)^{-1}]$.  Since $\Gamma$ decays exponentially in time, this conservative bound tracks our estimate \Eqref{eq:redundancy} very closely for all $\delta < 1/ (2 \ln 2)$.}. 


\begin{figure} [bt!]
  \centering
  \includegraphics[width=\widewidthfactor \columnwidth ]{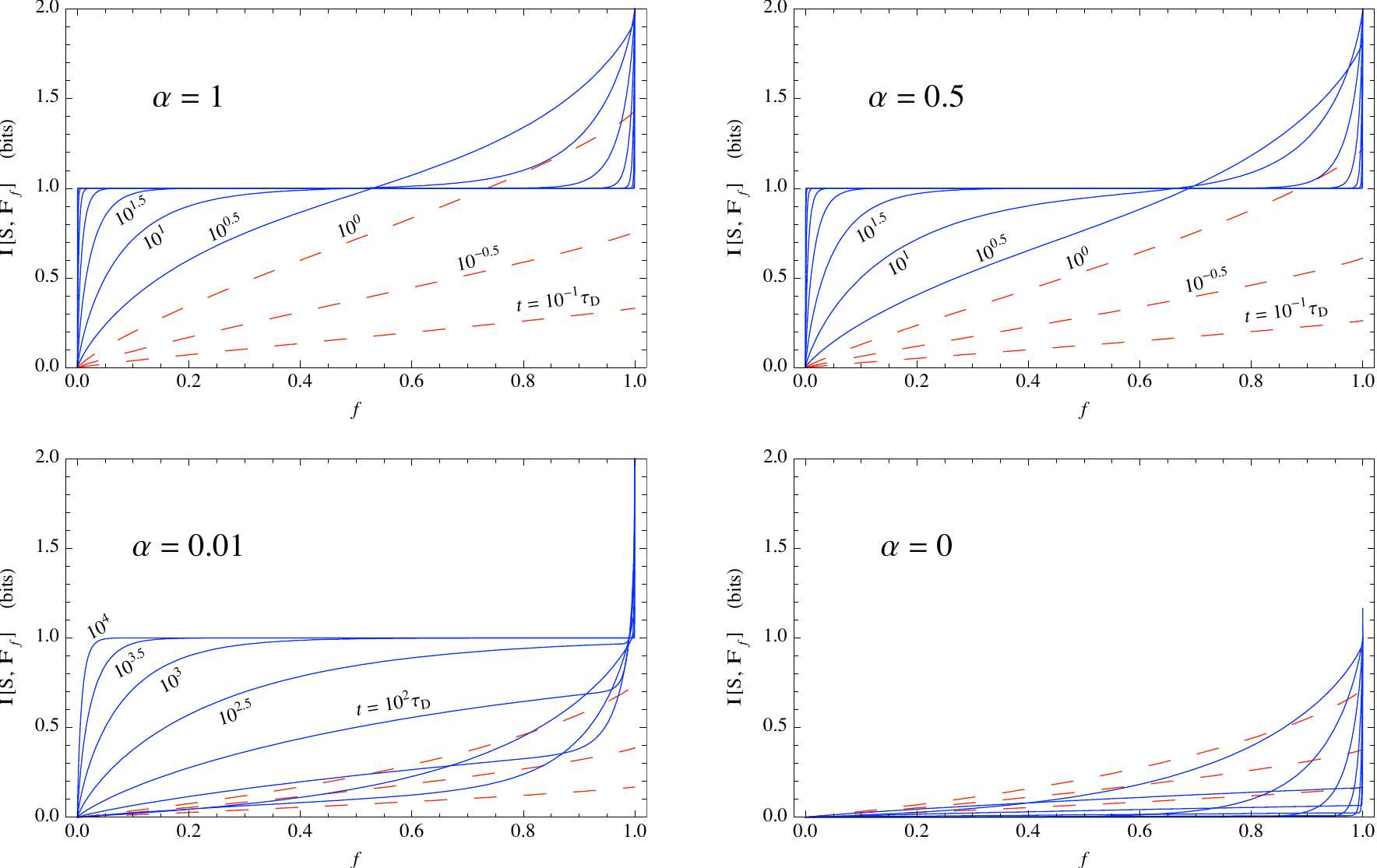}
  \caption{The quantum mutual information $\MI_{\Sys : \Frag_f}$ \Eqref{eq:universal-MI-open} versus fragment size $f$ at different elapsed times for an object illuminated by blackbody radiation.  Individual curves are labeled by the time $t$ in units of the decoherence time $\tau_D$ \Eqref{eq:universal-decoh-rate}. The same time slicing is used for all values of the receptivity $\alpha$.  Top left: $\alpha=1$.  For $t \le\tau_D$ (red dashed lines), the information about the system available in the environment is low.  The linearity in $f$ means each piece of the environment contains new, independent information.  For $t>\tau_D$ (blue solid lines), the plateau shape of the curve indicates redundancy; the first few pieces of the environment reveal a lot of information about the system, but additional pieces just confirm what is already known. On the plateau, the mutual information approaches it's maximum classical value, $\overline{H_\Sys} = 1$ bit $= \ln 2$ nats. The remaining information (i.e., above the plateau) is highly \emph{encoded} in the global state, in the sense that it can only be read by capturing almost all of $\Env$. Top right: $\alpha = 0.5$. Redundant copies are only produced at half the rate and the mutual information is no longer anti-symmetric about $f=0.5$ because the environment is mixed in the information-storing degrees of freedom.  Nevertheless, the plot approaches the same symmetric plateau shape for large $t$, illustrating that extensive redundancy is still achieved.  Bottom left: $\alpha = 0.01$. For receptivity this low (which is only expected for nearly isotropic illumination) the mutual information is initially greatly skewed.  Still, this just slows down acquisition of information by a factor of $\alpha^{-1} = 100$. (Compare the $t=10^{3.5} \tau_D$ curve for $\alpha = 0.01$ to the $t=10^{1.5} \tau_D$ for $\alpha = 1$.)  Bottom right: $\alpha = 0$.  Only for this idealized case of perfectly uniform illumination is information storage halted.  This is because the directional photon states are already ``full'' and cannot store more information about the state of the object.  Zero redundant copies are produced and the mutual information approaches $0$ as $t \to \infty$ for all $f < 1$.}
  \label{fig:all-alpha-QMI}
\end{figure}

The redundancy \Eqref{eq:redundancy} depends only weakly (logarithmically) on the information deficit $\delta$, which is consistent with previous results \cite{Zwolak2010, Blume-Kohout2006}.  The redundancy increases linearly with time at a rate proportional to the decoherence rate.  This is intuitive because (1) photons scatter off the object at a constant rate and (2) it is precisely the dependence of photon out states on the position of the object (roughly corresponding to a record) that causes decoherence.  

It can now be seen why we call $\alpha$ the receptivity of the environment; it determines, for a \emph{fixed} rate of decoherence, the rate at which records about the systems are created in the environment.  For maximum receptivity ($\alpha =1$), the  redundancy rate is equal to the decoherence rate and records are produced at the maximum speed.  For vanishing receptivity ($\alpha=0$), no redundant copies are produced no matter how effective the decoherence.

\subsection{Receptivity}

We now examine how the receptivity depends on the distribution of the illumination and, through the operator $\sonetwo{k}$, the scattering matrix.  However, most of our discussion will not rely on the particular angular dependence of the differential cross section.  

In general, both the decoherence rate $\tau_D^{-1}$ and the receptivity $\alpha$ will vary with different choices of $\bbb$.   The decoherence rate $\tau_D^{-1}$ changes for two reasons: (a) Different incoming photons contribute independently (that is, additively) to the decoherence rate, so larger or closer blackbodies covering more of the sky will naturally decohere the system faster.  (b)  Less dramatically, the contribution per unit solid angle to the rate of decoherence changes as the blackbody is moved around the unit sphere because of the factor of $3+11\cos^2 \theta$ in \Eqref{eq:key-matrix-element}.  The receptivity $\alpha$ similarly depends both on (a) the solid angle of the regions integrated over in \Eqref{eq:general-alpha}, and (b) the angular dependence of the integrand $|g(\hat{n},\hat{m})|^2$.

To disentangle these two quantities, simply consider a blackbody dimmed by some uniform intermediate medium.  Let $\beta \in [0,1]$ be the fractional degree to which the intensity of the illumination is reduced,  expressed by modifying the number density $N/V \to \beta N/V$ (equivalently, the intensity $I \to \beta I$).  The reduces the decoherence rate  accordingly, $\tau_D^{-1} \to \beta \tau_D^{-1}$, but leaves the receptivity $\alpha$ unchanged.  We can then consider an arbitrary family of blackbodies $\bbb^{(i)}$ with different $\alpha^{(i)}$, taking $i=0$ to be the one with the largest decoherence rate:  $\tau_{D}^{(0)\, -1} \geq \tau_{D}^{(i)\, -1}$ for all $i$.  If we dim each $\bbb^{(i)}$ by $\beta^{(i)} =  \tau_{D}^{(i) \, -1}/\tau_{D}^{(0)\, -1}$, we equalize all the decoherence rates without changing the receptivities.  This gives a physical interpretation for considering, with a fixed decoherence rate, how the receptivity (and hence the redundancy rate) depends on the shape of the blackbody.

In the appendix we show that $|g(\hat{n},\hat{m})|^2$, which is bounded, has angular dependence 
\begin{eqnarray}
\label{eq:g2}
|g(\hat{n},\hat{m})|^2 \propto (1+\cos^2 \theta_{n,m}) ( \cos \theta_{\Delta x, n} - \cos \theta_{\Delta x, m})^2 ,
\end{eqnarray}
where $\cos \theta_{a,b}=\hat{a} \cdot \hat{b}$ is the cosine of the angle between the unit vectors $\hat{a}$ and $\hat{b}$.  The explicit form of the receptivity for general $\bbb$ is then
\begin{eqnarray}
\label{eq:complete-alpha}
\alpha = \frac{\int_\bbb \dd \hat{n} \int_{\bbbb} \dd \hat{m}  \left(1 + \cos^2 \theta_{n,m} \right) \left( \cos \theta_{\Delta x, n} - \cos \theta_{\Delta x, m} \right)^2 }{\int_\bbb \dd \hat{n} \int_\sss \dd \hat{m} \left(1 + \cos^2 \theta_{n,m} \right) \left( \cos \theta_{\Delta x, n} - \cos \theta_{\Delta x, m} \right)^2} .
\end{eqnarray}
We know that $0 \le \alpha \le 1$ and, since $|g(\hat{n},\hat{m})|^2$ only vanishes when $\cos \theta_{\Delta x, n} = - \cos \theta_{\Delta x, m}$ (a set of measure zero on $\sss \times \sss$), we see that the two extremes are realized only in the following physical situations:

\begin{itemize}
\item For $\bbb = \{\hat{n}_0\}$ (point source illumination from the direction $ \hat{n}_0$), $\Omega = 0$, $\alpha = 1$, and we recover (13) of \cite{Riedel2010}:
\begin{eqnarray}
\label{eq:point-source-MI}
\MI_{\Sys : \Frag} = \ln 2 + \sum_{m=1}^{\infty} \frac{\Gamma^{(1-f) m} - \Gamma^{f m} - \Gamma^{m}}{2m(2m-1)} .
\end{eqnarray}
See \fref{fig:all-alpha-QMI}. Of course as noted above, exact point sources are a mathematical idealization; total flux of blackbody radiation is proportional to $\Omega$, so this is really the case when finite-size effects can be ignored.
 
\item For $\bbb = \sss$ (isotropic illumination), $\Omega = 4\pi$, $\alpha = 0$, and we recover (17) of \cite{Riedel2010}:
\begin{eqnarray}
\MI_{\Sys : \Frag_f} &= H_{\Sys d \Env} - H_{\Sys d \Env / \Frag} \\
\label{eq:isotropic-MI}
&= \sum_{n=1}^{\infty} \frac{\Gamma^{(1-f)n} -\Gamma^{n}}{2n(2n-1)} .
\end{eqnarray}
For times $t \gg \tau_D$, the entropy of the system as decohered by the entire environment, $H_{\Sys d \Env}$, and by just the complement of the fragment, $H_{\Sys d \Env /\Frag}$, are both exponentially close to $\ln 2$, so that $\MI_{\Sys : \Frag_f} = H_{\Sys d \Env} - H_{\Sys d \Env / \Frag}$ is only non-negligible for the brief period when $t \sim \tau_D$.  This is plotted in \fref{fig:all-alpha-QMI}, which shows that the mutual information barely rises from zero before fading away, never yielding a single redundant copy.  (For strictly vanishing $\alpha$, the approximation used for \Eqref{eq:redundancy} breaks down.)  The photon directional states, which are the component of the environment in which information about the object is stored, are initially fully mixed and so cannot hold any new information.  In other words, an observer relying on scattered radiation in an oven can see nothing \cite{Bennett2006}. This makes it clear that decoherence---which is \emph{maximized} for isotropic illumination---is not sufficient to guarantee redundancy \cite{Zurek2009}.
\end{itemize}

Between these two extremes, the form \Eqref{eq:general-alpha} indicates that the receptivity $\alpha$ will tend to decrease with increasingly large $\bbb$ \footnote{Even for a strictly enlarging sequence of blackbodies $\bbb(s)$ (where $\bbb(s_1) \subset \bbb(s_2)$ and $\Omega(s_1) < \Omega(s_2)$ for $s_1 < s_2$) the receptivity does not necessarily decrease monotonically with $\Omega$.  For the dipole scattering cross section considered in this work, as well as a few other example scattering operators, we have been able to construct unusual examples with $\bbb_1 \subset \bbb_2$ and $\Omega_1 < \Omega_2$ such that $\alpha_1 < \alpha_2$.  However, simple blackbody shapes like uniform disks have receptivity that decreases monotonically with solid angle $\Omega$.  The number of local extrema $\alpha$ takes with growing $\bbb$ is restricted by the size of higher frequency terms in the Fourier expansion of $|g(\hat{n},\hat{m})|^2$ and the complexity of the shape of $\bbb$.  Trivially, if $|g(\hat{n},\hat{m})|^2$ were constant, then the receptivity would just decrease linearly with $\Omega$.}. In physical situations (e.g., objects lit by light bulbs, the Sun, or ambient light), we expect illumination to be nonuniform.  This will correspond to receptivity that is not particularly close to either $0$ or $1$, so that the initial mixedness of the photon environment decreases the redundancy only by roughly a factor of order unity (in accordance with detailed calculations made of spin\nobreakdashes-$\frac{1}{2}$ systems \cite{Zwolak2010}).  Since even very tiny objects have extremely short decoherence times \cite{SchlosshauerText, Joos1985}, redundancies will still be very large for realistic illumination.

To get intuition about the dependency of $\alpha$ on $\bbb$, we again specialize to the case of the blackbody disk with solid angle $\Omega$ so that the integrals in \Eqref{eq:complete-alpha} can be preformed.  (See the appendix.)  The results for the decoherence rate, the receptivity, and redundancy rate have been plotted in \fref{fig:disk-alpha}.  



The receptivity of the decoherence is related to the haziness \cite{Zwolak2009a, Zwolak2010} (initial entropy) of the environment.  Environments with zero haziness in the information-storing degrees of freedom (photon directional states) will have maximum receptivity.  Environments with maximum haziness will have zero receptivity.  However, the receptivity is not a strict function of the haziness since it depends also on the form of the scattering operator, i.e. the way in which $\Sys$ makes its mark on $\Env$.


\newsavebox{\tallbox}
\newsavebox{\stretchbox}

\begin{figure} [bt!]

 \centering

\newcommand{\widener}{0.45}
\newcommand{\thinfyer}{0.47}
\newcommand{\muchthinfyer}{0.37}

\fboxsep=0pt
\sbox{\tallbox}{\includegraphics[width=\thinfyer \columnwidth]{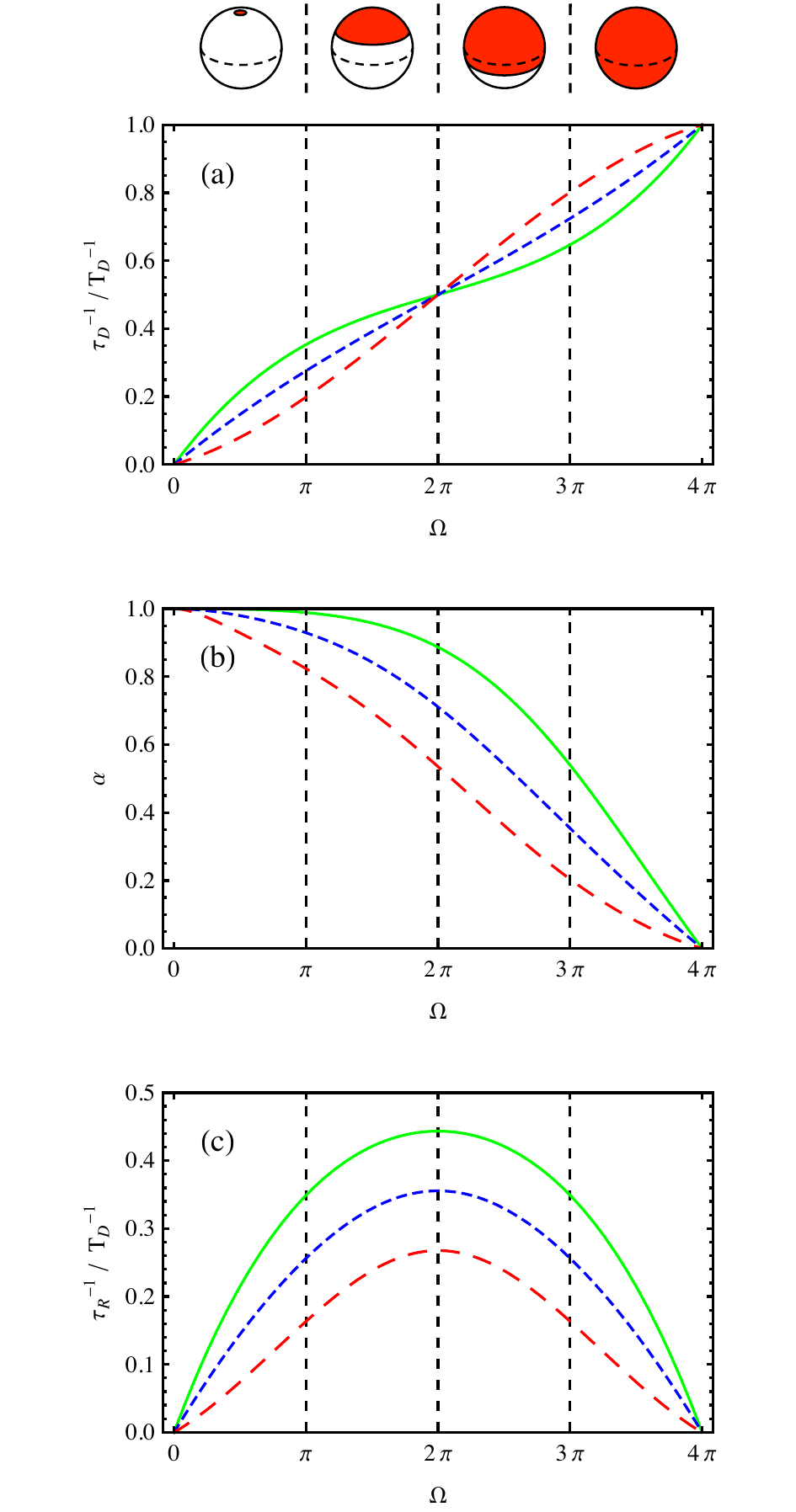}}
\sbox{\stretchbox}{\includegraphics[width=\widener \columnwidth]{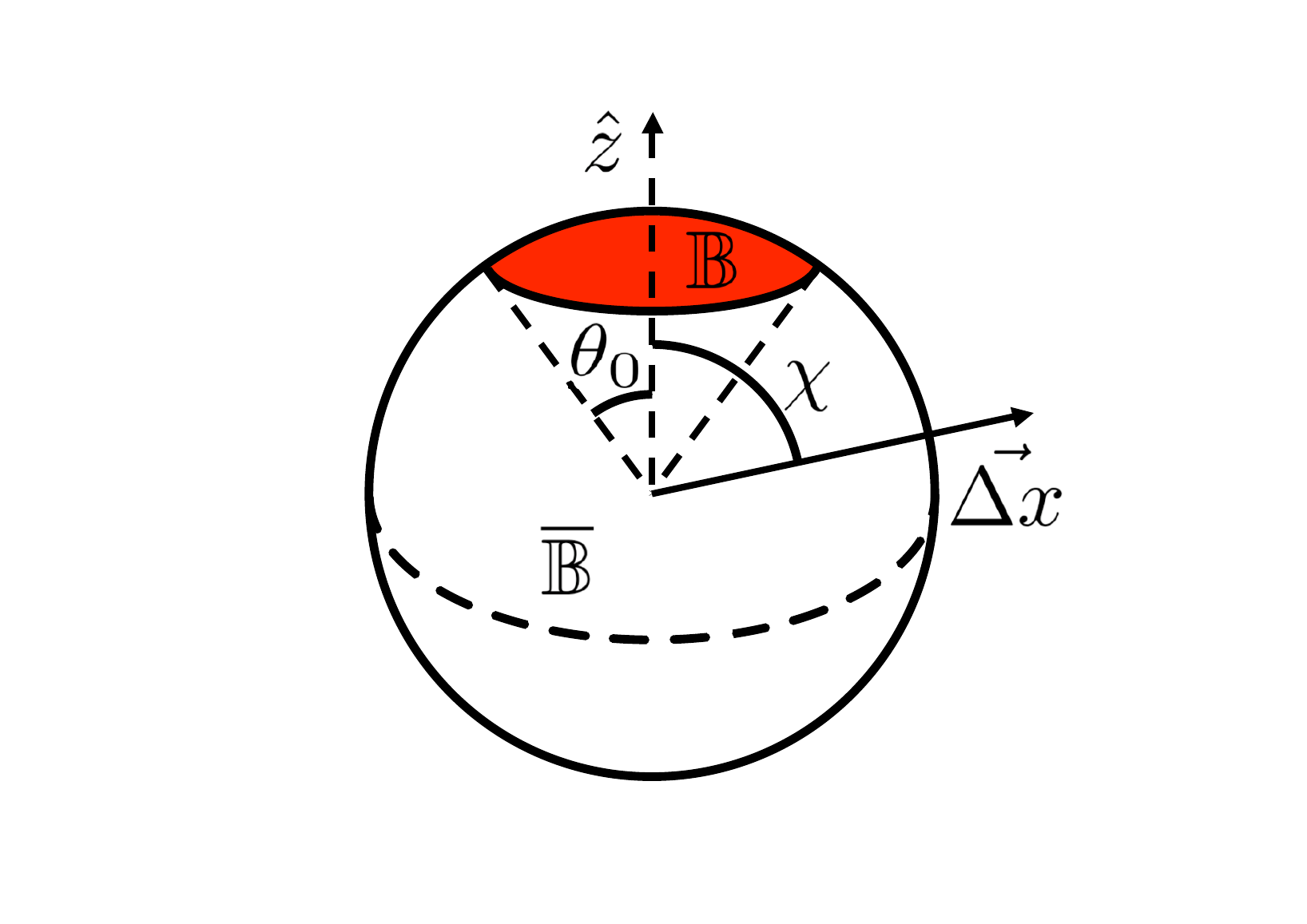}}

\subfigure{\vbox to \ht  \tallbox {  \vfil \hbox to \wd \stretchbox  {\includegraphics[width=\muchthinfyer \columnwidth]{disk_diagram_tight.pdf} }  \vfil  }}
\subfigure{\usebox{\tallbox}}

  \caption{  Decoherence and quantum Darwinism for an object illuminated by a blackbody disk.  Left: The disk has solid angle $\Omega$ and the center of the disk $\hat{z}$ makes an angle $\chi$ with $\vec{\Delta x}$. Precisely, $\bbb = \{ (\theta, \phi) \in \sss \mid \theta \le \theta_0 \}$ where $\cos \theta_0 = 1-\Omega/2\pi$.    Right: The decoherence rate, the receptivity, and the redundancy rate for a blackbody disk as a function of disk size.  We consider the three cases $\chi = 0^{\circ}$ (green solid line), $45^{\circ}$ (blue short-dashed line), and $90^{\circ}$ (red long-dashed line).  (a) The decoherence rate $\tau_D^{-1}$ \eqref{eq:norm-decoh-rate-disk} normalized by the maximum decoherence rate $T_D^{-1}$ obtained at $\Omega = 4 \pi$.  The decoherence rate is monotonically increasing with $\Omega$, a property that holds regardless of the scattering cross section.  (b)  The receptivity $\alpha$ \eqref{eq:alpha-disk} decreases monotonically with $\Omega$ (i.e. with increased angular mixing of the environment), a relationship that holds for arbitrary $\chi$.  (c) The redundancy rate $\tau_R^{-1} = \alpha \tau_D^{-1}$ normalized by $T_D^{-1}$.  The redundancy rate is symmetric about $\Omega = 2\pi$ because $\bbb$ and its complement $\bbbb$ appear on equal footing in \Eqref{eq:redun-prop}.  For $\Omega=0$, there is zero illumination and so there is neither decoherence nor quantum Darwinism.  For $\Omega = 4\pi$, the decoherence is maximized but there is no Darwinism because the receptivity vanishes; the angular mixing of the environment is total, allowing no recording of information. Away from either extreme the redundancy rate is within an order of magnitude of the decoherence rate.}
  \label{fig:disk-alpha}
\end{figure}


\subsection{Origin of difference between decoherence and redundancy rates}

The key difference between the redundancy rate $\tau_R^{-1}$ and the decoherence rate $\tau_D^{-1}$ is the type of matrix elements on which each depend.  Mere decoherence of the two pointer states $\ket{\vec{x}_1}$ and $\ket{\vec{x}_2}$ of $\Sys$ is determined only by the overlap of the $\Sys$-conditioned states of $\Env$, $\rho_{\Env}^{({\vec{x}_1})}$ and $\rho_{\Env}^{({\vec{x}_2})}$.  Initial mixedness of the environment is accounted for by simply averaging the overlap of the $\Sys$-conditioned pure states of $\Env$, $S_{\vec{x}_a}^{(k)} \ket{n}$, when computing the decoherence factor: 
\begin{eqnarray}
\label{eq:decoh-interp}
\gamma = \Abs{\operatorname{Tr} \left[ \sone \rho_{e}^0 \stwo^\dagger \right]}^2 = \Abs{ \int_0^{\infty} \dd k \, p(k)  \int_\bbb \frac{\dd \hat{n}}{\Omega} \matrixelement{\hat{n}}{\sonetwo{k}}{\hat{n}}}^2,
\end{eqnarray}
So long as this average overlap is unchanged, mixedness of the environment does not affect decoherence.  Importantly, the decoherence rate depends only on inner products between the \emph{same}  \footnote{The apparent dependence of \Eqref{eq:decoh-prop} on matrix elements of the form $\matrixelement{\hat{n}}{\sonetwo{k}}{\hat{m}}$ for $\hat{n} \neq \hat{m}$ can be removed by using the completeness relation $\int_\sss \dd \hat{m} \projector{\hat{m}} = I$.  The same cannot be done for \Eqref{eq:redun-prop}.} initial states of $\Env$ conditioned on different states of  $\Sys$.

\begin{figure} [bt!]
  \centering
  \includegraphics[width=\dualwidthfactor \columnwidth ]{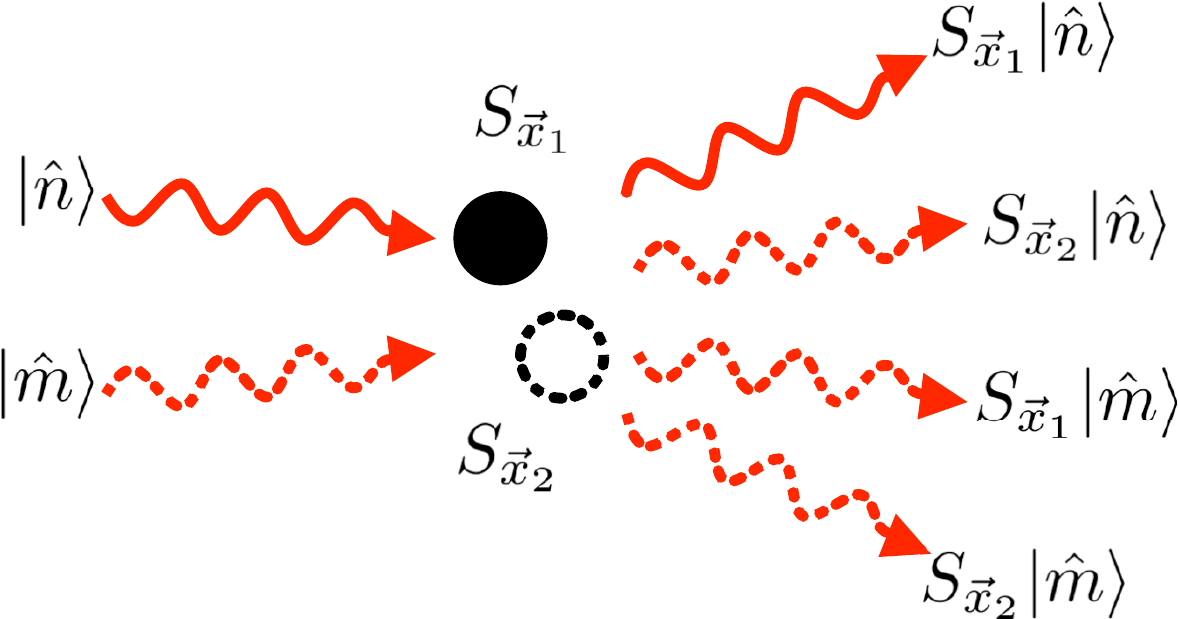}
  \caption{The various out states for different initial states of $\Env$ and different scattering states of $\Sys$.  Let $\hat{n}, \hat{m} \in \bbb$ be possible directions for incoming photons and let $\vec{x}_1$ and $\vec{x}_2$ be the pointer states of the object. (We ignore photon momentum, which decouples since the scattering is elastic.) Then there are four distinct out states and therefore four non-trivial inner products.  The decoherence rate $\tau_D^{-1}$ [see \Eqref{eq:decoh-prop}] depends only on inner products between \emph{like} initial states of the environment, $\matrixelement{\hat{n}}{\sonetwo{}}{\hat{n}}$ and $\matrixelement{\hat{m}}{\sonetwo{}}{\hat{m}}$, but the redundancy rate [see \Eqref{eq:redun-prop}] depends also on $\matrixelement{\hat{n}}{\sonetwo{}}{\hat{m}}$ and $\matrixelement{\hat{n}}{\stwoone{}}{\hat{m}}$. High redundancy depends not just on small overlap between the $\Sys$-conditioned states for the range of initial states of $\Env$ but also on the observer's ability to infer the state of $\Sys$ from imprints left on \emph{different} initial states of $\Env$.}
  \label{fig:redun-interp}
\end{figure}

On the other hand, the production of records is very sensitive to the initial mixedness.  The redundancy rate is proportional to the numerator of \Eqref{eq:discrete-alpha}.  That numerator includes matrix elements of the form $\Abs{\matrixelement{\hat{n}}{\sonetwo{k}}{\hat{m}}}^2$ where, crucially, $\hat{n} \neq \hat{m}$.  For redundant records to be produced, there must be small overlap between $\Sys$-conditioned states for \emph{different pure states of $\Env$ in the initial mixture}.  In other words, the observer must be able to distinguish the imprint of the pointer states of the system on different initial environment states.  When $\Abs{\matrixelement{\hat{n}}{\sonetwo{k}}{\hat{m}}}^2$ is not small for $\hat{n} \neq \hat{m}$, then the observer cannot tell whether she has sampled (a) a photon that started in state $\ket{\hat{n}}$ and scattered off the system in state $\ket{\vec{x}_1}$ or (b) a photon that started in state $\ket{\hat{m}}$ and scattered off the system in state $\ket{\vec{x}_2}$.  See \fref{fig:redun-interp}.
 

\section{General superpositions}
\label{sec:general}

So far we have considered only objects localized in a balanced ``cat'' state: $|\psi (\vec{x})|^2 \approx [\delta(\vec{x}-\vec{x}_1)+\delta(\vec{x}-\vec{x}_2)]/2$.  We now gather some evidence that relaxing this assumption to allow more general initial object states does not give qualitatively new behavior.

For an arbitrary initial wavefunction $\psi (\vec{x})$, the decoherence behavior is very simple.  There is no self-evolution of the object, and the initial off-diagonal terms $\rho_\Sys (\vec{x},\vec{x}')$ decay exponentially in time at a rate that is a function of $\Delta x = |\vec{x}-\vec{x}'|$.  As described in \cite{Gallis1990}, the rate is proportional to $\Delta x^2$ for small distances and saturates to a constant value for large distances.  We would like to know if the mutual information $\MI_{\Sys : \Frag}$ is similarly well-behaved for general initial object wavefunctions.

First, we consider the mutual information for the ``unbalanced cat'' state 
\begin{eqnarray}
\psi (\vec{x}) = \sqrt{p_1} \delta(\vec{x}-\vec{x}_1) + \sqrt{p_2} \delta(\vec{x}-\vec{x}_2) , \qquad p_1 + p_2 = 1,
\end{eqnarray}
illuminated by a point-source.  The new post-scattering density matrices are
\begin{eqnarray}
\rho_{\Sys \Env} &= \sum_{a,b = 1,2}  \sqrt{p_a p_b}  \ketbra{\vec{x}_a}{\vec{x}_b} \otimes \left(S_{\vec{x}_a} \rho_e^0 S_{\vec{x}_b}^\dagger \right)^{\bigotimes N} , \\
\rho_{\Sys} &= \sum_{a,b = 1,2}  \sqrt{p_a p_b}  \ketbra{\vec{x}_a}{\vec{x}_b} \left({\operatorname{Tr}}_e \left[S_{\vec{x}_a} \rho_e^0 S_{\vec{x}_b}^\dagger \right] \right)^{ N} , 
\end{eqnarray}
and
\begin{eqnarray}
\rho_{\Frag} = \sum_{a=1,2}  p_a  \left(S_{\vec{x}_a} \rho_e^0 S_{\vec{x}_a}^\dagger \right)^{\bigotimes fN}  = p_1 \left(\rho_e^{(\vec{x}_{1})}  \right)^{\bigotimes fN} +  p_2 \left(\rho_e^{(\vec{x}_{2})}  \right)^{\bigotimes fN}.
\end{eqnarray}
Diagonalizing $\rho_{\Sys}$ is not any more difficult than the balanced case, yielding the eigenvalues
\begin{eqnarray}
\lambda_\pm^\Sys &= \frac{1}{2} \pm \frac{1}{2} \sqrt{(p_1-p_2)^2 + 4 p_1 p_2  \Abs{\operatorname{Tr} \left[\sone \rho_e^0 \stwo^\dagger \right] }^{2 N} } \\
&= \frac{1}{2} \pm \frac{1}{2} \sqrt{\mu + (1-\mu)  \Gamma } ,
\end{eqnarray}
 where $\mu \equiv (p_1-p_2)^2$.  From this, we see $H_\Sys$ for the unbalanced cat can be obtained from the balanced case by making the replacement $\Gamma \to \mu + (1-\mu) \Gamma$.  (The decoherence factor, defined as the inner product between the environment states conditioned on pure system pointer states, is still $\Gamma$.)  

We decompose the photon Hilbert space as in the appendix, \Eqref{eq:decompose-frag}, only now with
\begin{eqnarray}
\rho_{\hat{\Frag}}^\chi &= p_1 \bigotimes_{i=1}^{f N} \left[ \sone^{k_i} |\hat{n}\rangle_i \langle \hat{n}| \sone^{k_i}{}^\dagger \right] + p_2 \bigotimes_{i=1}^{f N} \left[ \stwo^{k_i} |\hat{n}\rangle\langle \hat{n}| \stwo^{k_i}{}^\dagger \right] .
\end{eqnarray}
It is only a little more work to show that the eigenvalues of $\rho_{\hat{\Frag}}^\chi$ (which carry momentum dependence) are similarly modified as
\begin{eqnarray}
\lambda_\pm^\chi &= \frac{1}{2} \pm \frac{1}{2} \sqrt{\mu + (1-\mu) \prod_{i \in \Frag} \Abs{ \bra{n} \sonetwo{k_i} \ket{n} }^{2} } .
\end{eqnarray}
Carrying out the momentum integrals leads to the similar replacement $\Gamma^f \to \mu + (1-\mu) \Gamma^f$.  

\begin{figure} [bt!]
  \centering
  \includegraphics[width=\dualwidthfactor \columnwidth ]{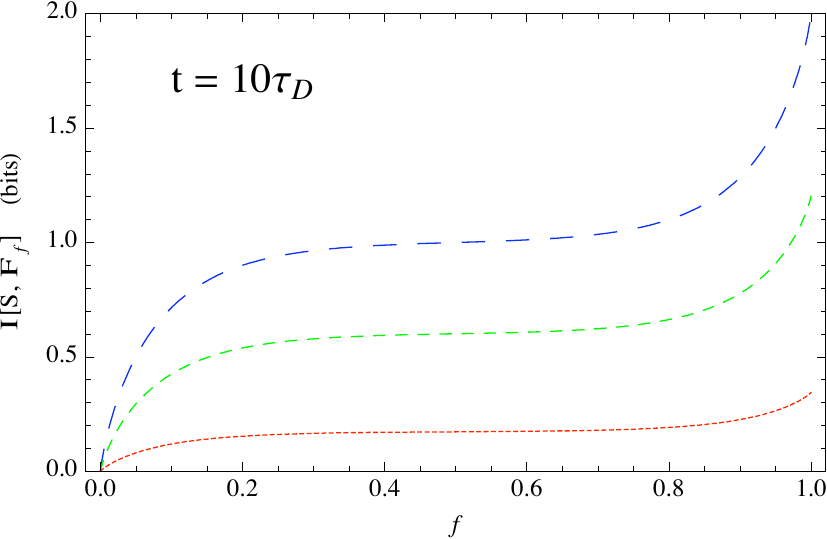}
  \includegraphics[width=\dualwidthfactor \columnwidth ]{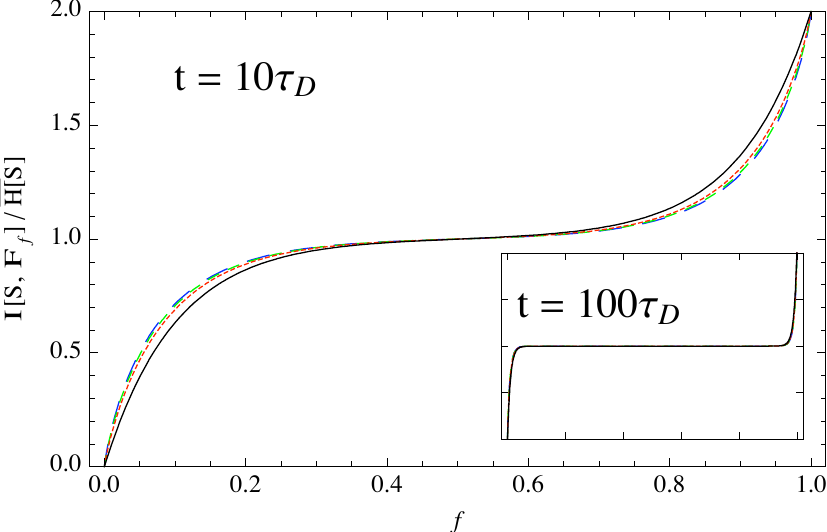}
  \caption{The partial information plot of unbalanced cat states, for $t = 10\, \tau_D$ and $\mu = (p_1 - p_2)^2 = 0$ (long-dashed blue line), $\mu = 0.5$ (medium-dashed green line), $\mu = 0.9$ (short-dashed red line), and $\mu \to 1$ (solid black line).  More unbalancing (increased $\mu$) moves the classical plateau down to the new maximum system entropy, $\overline{H_\Sys} = -p_1 \ln p_1-p_2 \ln p_2$, and softens the shoulders---but with only finite softening as $\mu \to 1$.  Left panel: unnormalized mutual information.  (The $\mu \to 1$ case is not visible.)  Right panel: renormalized mutual information, to clearly show softening of the shoulders. The inset shows the renormalized $t = 100\, \tau_D$ case.}
  \label{fig:unbalanced-QMI}
\end{figure}

The net effect of unbalancing the cat state on the mutual information is to make the replacement $x \to \mu + (1-\mu) x$, for $x = \Gamma$, $\Gamma^{f}$, or $\Gamma^{1-f}$, in \Eqref{eq:point-source-MI}.
The only qualitative change to the partial information plot is to lower the classical plateau to the new maximum system entropy, $\overline{H_\Sys} = -p_1 \ln p_1-p_2 \ln p_2 < \ln 2$, and to ``soften'' the shoulders.  (See \fref{fig:unbalanced-QMI}.) The mutual information for the extreme case ($\mu = 0$) is trivially zero, but the limit of the \emph{renormalized} mutual information for extremal probabilities is
\begin{eqnarray}
\lim_{\mu \to 1} \left(\frac{\MI_{\Sys : \Frag_f}} {\overline{H_\Sys}} \right)= 1 + \Gamma^{1-f} - \Gamma^{f } -\Gamma.
\end{eqnarray}
Note that there is finite softening as $\mu \to 1$.



\begin{figure} [bt!]
  \centering
  \includegraphics[width=\dualwidthfactor \columnwidth ]{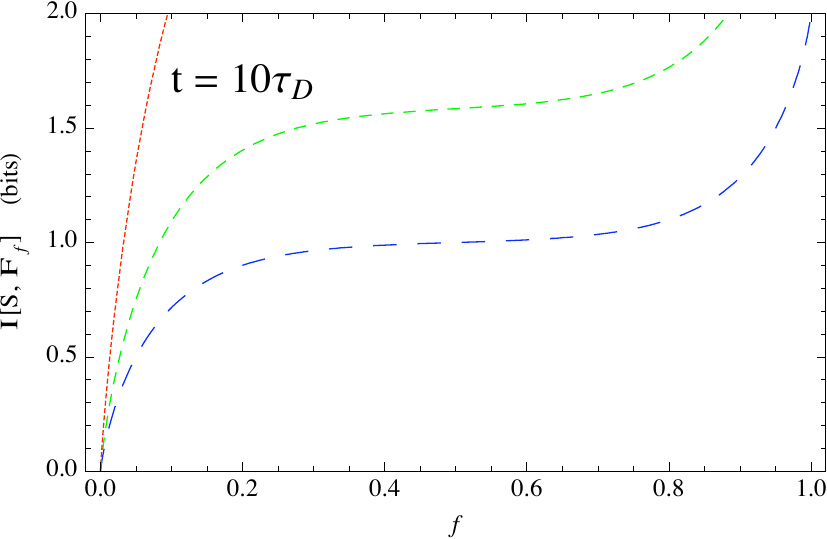}
  \includegraphics[width=\dualwidthfactor \columnwidth ]{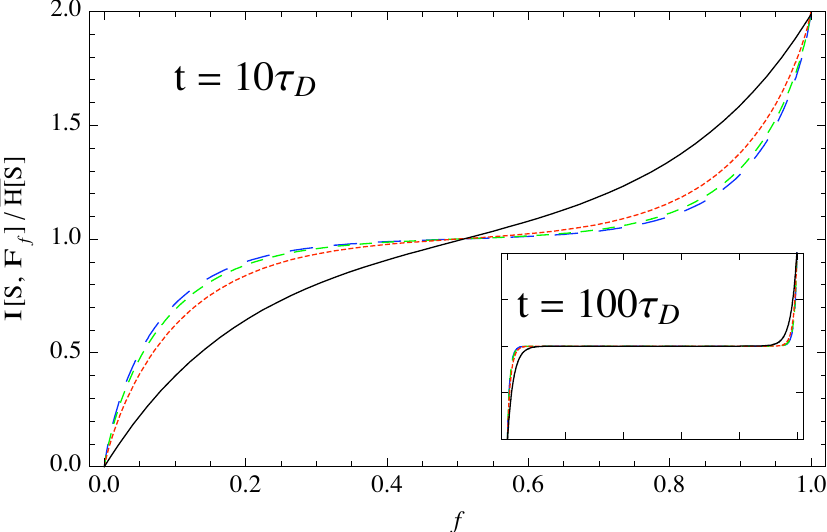}
  \caption{The partial information plot of  balanced $M$-way cat states for $t = 10\, \tau_D$ and $M=2$ (long-dashed blue line), $M=3$ (medium-dashed green line),  $M=10$ (short-dashed red line), and $M=\infty$ (solid black line).    Increasing $M$ moves the classical plateau to $\overline{H_\Sys} = \ln M$ and softens the shoulders, but with only finite softening as $M \to \infty$.  Left panel: unnormalized mutual information. (The $M=\infty$ case is not visible.)  Right panel: renormalized mutual information, to clearly show softening of the shoulders.  The inset shows the renormalized $t = 100\, \tau_D$ case.}
  \label{fig:mway-QMI}
\end{figure}


Now we look at a ``balanced $M$-way-cat'' state, $\psi (\vec{x}) = M^{-1/2} \sum_{a=1}^M \delta(\vec{x}-\vec{x}_a)$, exposed to a point-source blackbody.  Let us assume the special case where the decoherence factors for all off-diagonal elements of $\rho_\Sys$ are equal.  That is, assume $\bra{\hat{n}} S_{\vec{x}_a}{}^\dagger S_{\vec{x}_b}\ket{\hat{n}}$ is the same for all $a \neq b$ (such as when $M=3$ and $\vec{x}_1$, $\vec{x}_2$, and $\vec{x}_3$ form an equilateral triangle in a plane perpendicular to the direction of illumination).  
Generalizing from the original case of the balanced 2-way-cat does not require any new tricks.  The corresponding mutual information is
\begin{eqnarray}
\label{eq:mi-balanced-mway}
\fl \MI_{\Sys : \Frag_f} = \ln M + \frac{1}{M}\sum_{n=2}^{\infty} \frac{(M-1)+(1-M)^n}{n(n-1)} \left(\sqrt{\Gamma}^{(1-f)n} - \sqrt{\Gamma}^{f n} -\sqrt{\Gamma}^{n}\right) ,
\end{eqnarray}
which can be checked to reduce to \Eqref{eq:point-source-MI} for $M=2$.

This yields a partial information plot with slightly softer shoulders and, as expected, a classical plateau at $\overline{H_\Sys} = \ln M$.  (See \fref{fig:mway-QMI}.) The limit of the renormalized mutual information for large $M$ is
\begin{eqnarray}
\lim_{M \to \infty} \left(\frac{\MI_{\Sys : \Frag_f}}{\overline{H_\Sys}} \right)= 1 + \sqrt{\Gamma}^{1-f} - \sqrt{\Gamma}^{f } -\sqrt{\Gamma}.
\end{eqnarray}
Again, there is only finite softening for even the extreme case ($M \to \infty$).

Balanced $M$-way-cat states with arbitrary decoherence factors are difficult to handle because they require diagonalizing a matrix with $M (M-1)/2$ different off-diagonal terms.  We can still say the following.  Let $\gamma_{i,j}$, for $i \neq j$, be the set of decoherence factors.  Call the factor with the largest (smallest) absolute value $\gamma_{\mathrm{W}}$ ($\gamma_{\mathrm{S}}$), with ``W'' (``S'') standing for weak (strong) decoherence.  Let $\rho_\Sys^{\mathrm{W}}$ ($\rho_\Sys^{\mathrm{S}}$) be the hypothetical state resulting from setting all decoherence factors to $\gamma_{\mathrm{W}}$ ($\gamma_{\mathrm{S}}$), and likewise for $\rho_{\Sys \Frag}^{\mathrm{W}}$ ($\rho_{\Sys \Frag}^{\mathrm{S}}$) and $\rho_\Frag^{\mathrm{W}}$ ($\rho_\Frag^{\mathrm{S}}$).  

There is strong numerical evidence that we can bound $H_\Sys^{\mathrm{W}} \le H_\Sys \le H_\Sys^{\mathrm{S}}$, in agreement with intuition.  The same is true for $H_\Frag$ and $H_{\Sys \Frag}$ because, in the proper basis, $\rho_\Frag$ and $\rho_{\Sys \Frag}$ take the same form as $\rho_{\Sys}$.  Finally, we note that in the large-$t$ (small-$\Gamma$) limit, 
\begin{eqnarray}
0 < H_\Sys^{\mathrm{S}} - H_\Sys^{\mathrm{W}}&=&{} O(\Gamma_{\mathrm{W}}) \label{eq:strong-weak-H_S},
\\ 0 < H_\Frag^{\mathrm{S}} - H_\Frag^{\mathrm{W}} &=& {} O(\Gamma_{\mathrm{W}}^f) \label{eq:strong-weak-H_F},
\\ 0 < \, H_{\Sys \Frag}^{\mathrm{S}} - H_{\Sys \Frag}^{\mathrm{W}} &=&  O(\Gamma_{\mathrm{W}}^{1-f}) \label{eq:strong-weak-H_SF}.
\end{eqnarray}
This limit is quickly reached since $\Gamma = e^{-t/\tau_D}$, and this is exactly where we would like to calculate redundancy.  For times $t$ long enough to make higher order powers of $\Gamma$ negligible, the mutual information $\MI_{\Sys : \Frag_f}$---and hence the redundancy $R_\delta$---is bounded by the associated values for the strong and weak decoherence:
\begin{eqnarray}
\MI_{\Sys : \Frag_f}^{\mathrm{W}} <  \MI_{\Sys : \Frag_f} < \MI_{\Sys : \Frag_f}^{\mathrm{S}}, 
\\ R_\delta^{\mathrm{W}} <  R_\delta < R_\delta^{\mathrm{S}},
\end{eqnarray}
for $f < 0.5$.  In other words, having unequal decoherence factors for a balanced $M$-way cat state does not drastically alter the behavior of the mutual information; for times much larger than the decoherence time $\tau_D$, the redundancy can be bounded from above and below by calculations using the largest and smallest decoherence factor.

For ``unbalanced M-way-cat'' states, $\psi (\vec{x}) = \sum_{a=1}^M p_a \delta(\vec{x}-\vec{x}_a)$, (\ref{eq:strong-weak-H_S} - \ref{eq:strong-weak-H_SF}) still hold, but we cannot use \Eqref{eq:mi-balanced-mway} to get a power series for the entropies in terms of $\Gamma$.  But we do know that the classical plateau must still form at $\overline{H_\Sys} = - \sum_{a=1}^M p_a \ln p_a$ since, for large times, all off-diagonal terms are driven to zero.  As $M \to \infty$, these states can approximate a generic continuous wavefunction, but subtleties enter.  The maximum entropy $\overline{H_\Sys} = \ln M$ diverges, while the decoherence factor associated with two sufficiently nearby points becomes arbitrarily weak.  For these small distances, the self-Hamiltonian of the system will no longer be negligible on the times scales of decoherence.

\newtext
\section{Conclusion}
\label{sec:conclusion}

We illustrate quantum Darwinism \cite{Zurek2000, Ollivier2005, Zurek2009, Zurek2007b} in systems decohered by blackbody radiation.  Such radiation is by far the dominant form of illumination in everyday life and is the medium though which we, as observers, gather most of our information.  The huge redundancy growth rates we have calculated support the claim that a purely quantum universe can account for the appearance of objective and robust classical states.  

This is the first realistic model of quantum Darwinism, and the first with an environment with the capacity for two distinct types of mixing.  The most important type of mixing is in that component of the environment responsible for storing information about the system (the angular degrees of freedom).  We have shown, in agreement with previous studies of abstract systems \cite{Zwolak2009a,Zwolak2010}, that mixing of this type decreases the environment's ability to record information about the system---\emph{without} decreasing its ability to decohere.  The other type of mixing (the energy spectrum) only effects records insofar as certain modes (higher energy) are better able to resolve, and therefore record, states of the system.

We have extended the model to more general initial states, giving evidence that the qualitative features of the mutual information and redundancy are robust.

\newtext
\ack

We thank Robin Blume-Kohout, Haitao Quan, and Michael Zwolak for helpful discussions. One of us (CJR) is grateful to Steve Flammia, Helge Kr\"uger, and Willie Wong for mathematical insight. This research is supported by the U.S. Department of Energy through the LANL/LDRD program and, in part, by the Foundational Questions Institute (FQXi) and the John Templeton Foundation.



\appendix

\section*{Appendix}
\setcounter{section}{1} 

In this appendix we calculate $H_\Frag$ \Eqref{eq:delta-H_F}, the entropy of a fragment $\Frag$ of the photon environment $\Env$.  We take advantage of the special form of our model; the elastic scattering conserves energy but mixes photon direction.  This alows $\rho_{\Frag}$ to be decomposed into $k$-blocks:
\begin{eqnarray}
\label{eq:decompose-frag}
\rho_{\Frag}  = \int \dd \chi \, p(\chi) \, \projector{\chi} \otimes \rho_{\hat{\Frag}}^\chi \, \, ,
\end{eqnarray}
where $\chi = (k_1,\ldots, k_{f N})$ is the vector of the magnitudes of the photon momenta of \Frag, $p(\chi) = \prod_{i=1}^{f N} p(k_i) $ is the momentum spectrum probability distribution, and $\projector{\chi} = \bigotimes_{i \in \Frag} \projector{k_i}$.  The normalized $\rho_{\hat{\Frag}}^\chi$, which lives in the Hilbert space $\hat{\Frag}$ of angular eigenstates, is the density matrix conditional on the particular set of momenta $\chi$.

Since $\rho_\Frag$ is block-diagonal, it's entropy is
\begin{eqnarray}
\label{eq:H_F}
H_{\Frag} &=&  f N H[p(k)] + \int \dd \chi \, p(\chi) \, H_{\hat{\Frag}}^{\chi} \, ,
\end{eqnarray}
where $H_{\hat{\Frag}}^{\chi}$ is the entropy of $\rho_{\hat{\Frag}}^{\chi}$ and $H[p(k)] = H[p(\chi)] / f N$ is the entropy associated with the energy distribution $p(k)$ of a single thermal photon (which diverges since the photon Hilbert space is infinite dimensional).


Handling the mutual information is easier if we discretize the angular directions. Let $\Delta \Omega$ be the solid angle associated with a single discretized state and $D_\bbb = \Omega/\Delta \Omega$ be the dimension of the projector onto the directions in $\bbb$.  Then send
\begin{eqnarray}
\int \dd \hat{n} \to \sum_{\hat{n} \in \bbb} \Delta \Omega, \qquad  \ket{\hat{n}} \to \frac{1}{\sqrt{\Delta \Omega}} \ket{n}.
\end{eqnarray}

In the discrete picture, the conditional angular states are
\begin{eqnarray}
\label{eq:disk-source-decompose-frag}
\rho_{\hat{\Frag}}^\chi = \frac{1}{2} \left[ \bigotimes_{i \in \Frag} \frac{P^{(i)}}{D_\bbb} + \bigotimes_{i \in \Frag} \frac{Q^{(i)}}{D_\bbb} \right] = \frac{P + Q}{2 D_\bbb^{fN}} , 
\end{eqnarray}
where
\begin{eqnarray}
P = \bigotimes_{i \in \Frag} P^{(i)} , \qquad Q = \bigotimes_{i \in \Frag} Q^{(i)} ,
\\ \label{eq:P-i}
P^{(i)} =  \sum_{\hat{n} \in \bbb} \sone^{k_i} |n\rangle_i \langle n| \sone^{k_i} {}^ \dagger , 
\\ \label{eq:Q-i}
Q^{(i)} =  \sum_{\hat{n} \in \bbb} \stwo^{k_i} |n\rangle_i \langle n| \stwo^{k_i} {}^ \dagger . 
\end{eqnarray}
Notice that $P^{(i)}$ and $Q^{(i)}$ are $D_\bbb$-dimensional projectors unitarily related by $U^{(i)} = \stwo^{k_i} \sone^{k_i}{}^\dagger$.  $P$ and $Q$ are likewise unitarily equivalent and therefore they can be written in an appropriate basis as \cite{Halmos1969} \footnote{There is a technical complication that subspaces on which $P$ and $Q$ commute (which necessarily exist when $D_\bbb$ is odd) must be handled separately.  This is straightforward, with no changes to the final results.}
\begin{equation}
\label{eq:generic-position}
P = \left( \begin{array}{cc} I & 0 \\ 0 & 0 \end{array} \right)  ,  \qquad Q = \left( \begin{array}{cc} C^2 & C S \\ C S & S^2 \end{array} \right) ,
\end{equation}
where $C$ and $S$ are commuting positive matrices obeying $C^2 + S^2 = I$.  (Their eigenvalues are the cosines and sines of the canonical angles between the subspaces into which $P$ and $Q$ project.)  The eigenvalues of $\rho_{\hat{\Frag}}^\chi = (P+Q)/2 D_\bbb^{fN}$ are given by $(1 \pm |C|)/2 D_\bbb^{fN}$.  To calculate the $C$, it turns out that it is easier to diagonalize the related matrix
\begin{eqnarray}
PQP = \bigotimes_{i \in \Frag} P^{(i)} Q^{(i)} P^{(i)} = \bigotimes_{i \in \Frag} \left( \begin{array}{cc} (C^{(i)})^2 & 0 \\ 0 & 0 \end{array} \right) =  \left( \begin{array}{cc} C^2 & 0 \\ 0 & 0 \end{array} \right)   .
\end{eqnarray}
By \Eqref{eq:P-i} and \Eqref{eq:generic-position}, the first subspace corresponds to the span of the basis $\{\ket{\tilde{n}} \equiv \sone^{k_i} \ket{n} \mid n \in \bbb\}$. The elements of $P^{(i)} Q^{(i)} P^{(i)}$ in that subspace are given by
\begin{eqnarray}
\bra{\tilde{n}} P^{(i)}  Q^{(i)} P^{(i)} \ket{\tilde{n}'} &=& \sum_{\hat{m} \in \bbb} \bra{n} \sonetwo{k_i} \ket{m} \bra{m} \stwoone{k_i} \ket{n'} 
\\ \label{eq:define-B}
&=& \sum_{\hat{m} \in \bbb}  \braket{n}{m} \braket{m}{n'}  + \bra{n} B \ket{n'} 
\\ &=& \delta_{n, n'}  + B_{n, n'} .
\end{eqnarray}
By \Eqref{eq:key-matrix-element}, the elements of the Hermitian matrix $B$ defined above are of order $V^{-1}$.  Call the eigenvalues ${b(j)}$, with $j=1,\dots,D_\bbb$.  The eigenvalues of $\rho_{\hat{\Frag}}^\chi$ are then
\begin{eqnarray}
\lambda_{\vec{J},\pm} = \frac{1}{2 D_\bbb^{fN}} \left[ 1 \pm \prod_{i \in \Frag} \sqrt{1+b(j_i)} \right] ,
\end{eqnarray}
which are indexed by the vector $\vec{J} = (j_1,\dots,j_{fN})$ and the sign $\pm$.  When we plug these into the entropy formula and Taylor expand the logarithms, we get
\begin{eqnarray}
\fl \Delta H_{\hat{\Frag}}^\chi &=& H_{\hat{\Frag}}^\chi - f N \ln D_\bbb 
\\ \fl &=& \ln 2 - \frac {1}{2 D_\bbb^{f N}} \sum_{\vec{J}} \sum_{\pm} \left[ 1 \pm \prod_{i \in \Frag} \sqrt{1+b(j_i)} \right] \ln \left[ 1 \pm \prod_{i \in \Frag} \sqrt{1+b(j_i)} \right]
\\ \fl &=& \ln 2 - \frac {1}{D_\bbb^{f N}} \sum_{\vec{J}} \sum_{m=1}^{\infty} \frac{\prod_{i \in \Frag} \left[1+b(j_i)\right]^m}{2m(2m-1)}
\\ \fl &=& \ln 2 - \frac {1}{D_\bbb^{f N}} \sum_{m=1}^{\infty} \frac{1}{2m(2m-1)} \prod_{i \in \Frag} \left[\sum_{s=0}^m \binom{m}{s} \sum_{j_i=1}^{D_\bbb} b(j_i)^s \right]
\end{eqnarray}
Because the $b(j)$ are of order $V$, we need only keep the $s=0,1$ terms in the sum over $s$.  It then makes sense to define
\begin{eqnarray}
z(k) &\equiv& \frac{1}{D_\bbb} \sum_{j=1}^{D_{\bbb}} b(j)
\\ &=& \frac{1}{D_\bbb}  \Tr \left[P^{(i)}Q^{(i)}P^{(i)} - \sum_{\hat{m} \in \bbb} \projector{m} \right]
\\ \label{eq:z} &=& \frac{1}{D_\bbb}  \sum_{n,m \in \bbb} \left|\bra{n} \sonetwo{k} \ket{m}\right|^2 - 1
\end{eqnarray}
We now integrate over $\chi$ to get the full entropy for $\rho_{\hat{\Frag}}$:
\begin{eqnarray}
\fl \Delta H_{\hat{\Frag}} &=& \int \dd \chi p(\chi) \Delta H_{\hat{\Frag}}^\chi
\\ \fl &=& \ln 2 - \sum_{m=1}^{\infty} \frac{1}{2m(2m-1)} \int \dd \chi \, p(\chi) \prod_{i \in \Frag} \left[1 + \frac{m}{D_\bbb} \sum_{j_i=1}^\infty b(j_i) + O(V^{-2}) \right]
\\ \fl &=& \ln 2 - \sum_{m=1}^{\infty} \frac{1}{2m(2m-1)} \left[1 + m \int_0^\infty \dd k \, p(k) z(k) + O(V^{-2}) \right]^{f N}
\\ \fl &=& \ln 2 - \sum_{m=1}^{\infty} \frac{1}{2m(2m-1)} \left[ \exp \left( m Z f N \right) + O(V^{-2}) \right]
\end{eqnarray}
with $Z \equiv \int_0^\infty \dd k \, p(k) z(k)$.    Then we define
\begin{eqnarray}
\alpha &\equiv& \frac{Z }{\ln \gamma} 
\\ \label{eq:discrete-alpha}
&=& \frac{ D_\bbb^{-1} \int_0^\infty \dd k \, p(k) \sum_{\hat{n},\hat{m} \in \bbb} \left|\bra{n} \sonetwo{k} \ket{m}\right|^2 - 1} {\ln \left| \int_0^\infty \dd k \, p(k) \sum_{\hat{n} \in \bbb} D_\bbb^{-1} \bra{n} \sonetwo{k} \ket{n}\right|^2 }
\end{eqnarray}
so that 
\begin{eqnarray}
\exp(m Z f N) = \Gamma^{m \alpha f} = e^{-m \alpha f t/ \tau_D} . 
\end{eqnarray}
Moving to the $N,V \to \infty$ limit, we get
\begin{eqnarray}
\Delta H_{\hat{\Frag}} = H_{\hat{\Frag}} - H_{\hat{\Frag}}^0 =  \ln 2 - \sum_{m=1}^{\infty} \frac{\Gamma^{m \alpha f}}{2m(2m-1)}  
\end{eqnarray}
and
\begin{eqnarray}
\label{eq:continuous-alpha-A}
\alpha = \frac{\int_\bbb \dd \hat{n} \int_{\bbbb} \dd \hat{m}  |g(\hat{n},\hat{m})|^2 }{\int_\bbb \dd \hat{n} \int_\sss \dd \hat{m} |g(\hat{n},\hat{m})|^2} ,
\end{eqnarray}
where $\Abs{g(\hat{n},\hat{m})}^2 \equiv \int \dd k p(k) \Abs{\matrixelement{\hat{n}}{(\sonetwo{k}-I)}{\hat{m}}}^2$.  Note that $|g(\hat{n},\hat{m})|^2$ is symmetric in $\hat{n}$ and $\hat{m}$ for any scattering operator that is rotationally invariant.


In passing, we note that, from \Eqref{eq:general-alpha},
\begin{eqnarray}
\label{eq:decoh-prop}
\tau_D^{-1} \propto \int_\bbb \dd \hat{n} \int_\sss \dd \hat{m} |g(\hat{n},\hat{m})|^2
\end{eqnarray}
and
\begin{eqnarray}
\label{eq:redun-prop}
\tau_R^{-1} \propto \int_\bbb \dd \hat{n} \int_\bbbb \dd \hat{m} |g(\hat{n},\hat{m})|^2 .
\end{eqnarray}
Allowing bars to denote quantities associated with the complementary blackbody $\bbbb$, this means that the two decoherence rates sum to that of isotropic illumination,
\begin{eqnarray}
\tau_D^{-1} + \bar{\tau}_D^{-1} = T_D^{-1} ,
\end{eqnarray}
and that the redundancy rate is invariant when interchanging $\bbb \leftrightarrow \bbbb$ ,
\begin{eqnarray}
\tau_R^{-1} = \alpha / \tau_D = \bar{\alpha} / \bar{\tau}_D = \bar{\tau}_R^{-1} .
\end{eqnarray}

Previous studies of collisional decoherence \cite{Joos1985,Diosi1995,Gallis1990,Hornberger2003a,Hornberger2006} required the calculation of $\matrixelement{\hat{n}}{\sonetwo{k}}{\hat{m}}$ only for $\hat{n} = \hat{m}$.  We will use most of the same tricks to handle the case of $\hat{n} \neq \hat{m}$:
\begin{eqnarray}
\fl |\matrixelement{\hat{n}}{\sonetwo{k}}{\hat{m}}|^2 &=& |\matrixelement{\hat{n}}{(I- i T_k+\cdots) e^{i \vec{\Delta x} \cdot k  \hat{P} / \hbar}(I+i T_k+\cdots)}{\hat{m}}|^2
\\ \fl &=&  |\matrixelement{\hat{n}}{T_k}{\hat{m}}|^2 + |\matrixelement{\hat{n}}{T_k}{\hat{m}}|^2 
\\ \fl \nonumber && \qquad - \left(\matrixelement{\hat{n}}{T_k}{\hat{m}}\matrixelement{\hat{m}}{T_k}{\hat{n}} e^{i \vec{\Delta x} \cdot k (\hat{n} - \hat{m})/\hbar} + \mathrm{c.c.} \right) +\cdots \\ \fl &=&   2 |\matrixelement{\hat{n}}{T_k}{\hat{m}}|^2 \left[ 1 - \cos \left(\vec{\Delta x} \cdot k (\hat{n} - \hat{m})/\hbar \right) \right] + \cdots,
\end{eqnarray}
so
\begin{eqnarray}
\label{eq:exact-g2}
\fl |g(\hat{n},\hat{m})|^2 &=& \int \dd k p(k) |\matrixelement{\hat{n}}{\sonetwo{k}}{\hat{m}}|^2
\\ \fl &=& \int \dd k p(k) \left[ \frac{k^6 \tilde{a}^6 }{8 \pi^2 \hbar^8} \left(1 + \cos^2 \theta_{n,m} \right) \left[ \vec{\Delta x} \cdot k (\hat{n} - \hat{m})/\hbar)\right]^2 + \cdots \right]
\\ \fl &\propto& \left(1 + \cos^2 \theta_{n,m} \right) \left( \cos \theta_{\Delta x, n} - \cos \theta_{\Delta x, m} \right)^2 +  \cdots ,
\end{eqnarray}
Above, $T_k$ is the Hermitian operator that generates the unitary $S^k_{\vec{x}=0}$ for scattering off an object located at the origin, $\vec{x}=0$, and $\hat{P}$ is the momentum direction operator.  Ellipses denote higher-order terms ignored in the dipole ($\lambda \gg a$) and short-separation ($\lambda \gg \Delta x$) limits.  In the last line we have isolated the angular dependence since all other factors will cancel in the formula, \Eqref{eq:continuous-alpha-A}, for $\alpha$:
\begin{eqnarray}
\label{eq:complete-alpha-A}
\alpha = \frac{\int_\bbb \dd \hat{n} \int_{\bbbb} \dd \hat{m}  \left(1 + \cos^2 \theta_{n,m} \right) \left( \cos \theta_{\Delta x, n} - \cos \theta_{\Delta x, m} \right)^2 }{\int_\bbb \dd \hat{n} \int_\sss \dd \hat{m} \left(1 + \cos^2 \theta_{n,m} \right) \left( \cos \theta_{\Delta x, n} - \cos \theta_{\Delta x, m} \right)^2} .
\end{eqnarray}
This formula allows one to calculate the exact receptivity for any blackbody $\bbb$ for which the integrals can be performed.  

If we specialize to the case of a blackbody disk, $\bbb = \{ (\theta, \phi) \in \sss \mid \theta \le \theta_0\}$, whose center makes an angle $\chi$ with $\vec{\Delta x}$, then
\begin{eqnarray}%
\label{eq:alpha-disk}
\fl \alpha &=  \left[-117 c_\theta^6+295 c_\theta^4-575 c_\theta^2+685+ 6 c_\chi^2 \left(21 c_\theta^6-55 c_\theta^4+135 c_\theta^2+75 \right)\right] \times   
\\ \fl  \nonumber & \qquad   (c_\theta+1)  \left[32\left(40+11   c_\theta (1+c_\theta)  \left(3 c_\chi^2 -1\right) \right) \right]^{-1}
\end{eqnarray}
where we have abbreviated $c_\theta \equiv \cos \theta_0$, $c_\chi \equiv \cos \chi$.  This can be combined with the previously calculated decoherence rate  \eqref{eq:norm-decoh-rate-disk} to get redundancy rate $\tau_R^{-1} = \alpha \, \tau_d^{-1}$ for the disk.  All three quantities are plotted in \Figref{fig:disk-alpha} in terms of the solid angle $\Omega = 2 \pi(1-\cos \theta_0)$.

\bibliographystyle{ieeetr}
\bibliography{riedelbib}

\end{document}